\documentclass[twocolumn,showpacs,preprintnumbers,amsmath,amssymb]{revtex4}
\usepackage{graphicx}
\usepackage{amssymb}
\usepackage{amsmath}
\usepackage{dcolumn}
\usepackage{bm}

\begin{document}
\widetext 

\title{Flow-correlated dilution of a regular network leads to a percolating
  network during tumor induced angiogenesis}

\thispagestyle{empty}

\author{R. Paul}
\affiliation{Department of Neurobiology, Physiology and Behavior, University of California, Davis, CA 95616.}
\date{today}
\begin{abstract}
  We study a simplified stochastic model for the vascularization of a
  growing tumor, incorporating the formation of new blood vessels at
  the tumor periphery as well as their regression in the tumor center.
  The resulting morphology of the tumor vasculature differs
  drastically from the original one. We demonstrate that the
  probabilistic vessel collapse has to be correlated with the blood
  shear force in order to yield percolating network structures. The resulting
  tumor vasculature displays fractal properties. Fractal dimension,
  microvascular density (MVD), blood flow and shear force has been
  computed for a wide range of parameters.
\end{abstract}


\maketitle  
\section{Introduction}
A tumor growing beyond a critical size can only do so
by remodeling its surrounding blood vessel network that supply
nutrients and oxygen and removes waste products. This process is
initiated by tumor cells secreting various growth factors that induce
angiogenesis, the formation of new blood vessels from existing ones
\cite{Carmeliet00,Sansone01,Hanahan96,Maisonpierre97}. The neo-vascularization
primarily occurs in the periphery of the tumor which gives rise to a peripheral region with
substantially increased Micro-Vascular Density (MVD)~\cite{Holash99,Doeme02}. Inside the tumor
the MVD is usually drastically decreased, often leading to a necrotic
core.
  
Previous studies by G\"odde {\it et al}~\cite{Goedde01} focused on the
angiogenesis showed that blood shear stress-dependent rather than
pressure-dependent growth leads to a homogeneous distribution of
capillaries. However, the influence of a growing 
tumor on the angiogenic vasculature was missing. Several other studies have
focused on the angiogenesis in presence of a static
tumor~\cite{Anderson98,Levine01} or growth of a tumor within fixed
vasculature~\cite{Alarcon03}. More recently biologically motivated hybrid
cellular automaton models have been introduced by Bartha {\it et al}, Lee {\it
  et al} and Welter {\it et al}~\cite{Bartha06,Lee06,Welter07,Welter09} to
study this remodeling process of the tumor vasculature via cooption, vessel growth and
regression. Here the vasculature is
modeled by a network of pipes carrying a flow that 
is a source for a oxygen or nutrient concentration field, whereas the tumor is modeled as a
growth-and-death process that involves discrete cells proliferating or
dying in dependence of the local level of this concentration
field. Simultaneously the tumor cells are the sources for a growth
factor concentration field that triggers the formation of new links in
the vessel network at the tumor periphery and the dilation of pipes in
the tumor's interior. Vessels can also regress inside the tumor due
to increased solid stress or high acidity. Their models reproduce many known
features of experimentally analyzed tumor vasculature like the
compartmentalization into a highly vascularized perimeter, a well perfused
tumor periphery and necrotic core with only a few thick vessels threading it
\cite{Carmeliet00,Doeme02}.

In addition to the inhomogeneous MVD, tumor vasculatures are abnormal in various
ways, e.g., leaky, tortuous and dilated, have excessive branching and undergo
constant regression and remodeling~\cite{Maisonpierre97}. In particular it
turned out that the emerging vasculature has fractal properties reminiscent of
percolation clusters~\cite{Gazit98,Baish98,Baish00,Baish01}. In
\cite{Bartha06,Lee06,Welter07} it was argued that these specific geometric 
properties were the consequences of a flow correlated percolation process.

In order to disentangle the basic mechanism responsible for this
process from the various biological details of the model described in
\cite{Bartha06,Lee06,Welter07} we formulate here a drastically
simplified version of it, comprising, as we propose, the essential features
leading to the global characteristics of the vessel network morphology. In
essence we retain only the underlying dynamical pipe network capable of carrying a flow plus a
circular tumor that increases its radius linearly in time. This circular
region is further subdivided into a peripheral annulus where new pipes are 
inserted into the vessel network, and an inner annulus, where pipes
are dilated and/or removed from the network. The latter process turn
out to produce the flow correlated percolation process observed in
\cite{Bartha06,Lee06,Welter07}.

The organization of the paper is as follows: In section~\ref{def_model} the model
is defined, which represents a stochastic process that is studied with
Monte Carlo simulation. The results of these simulations are presented
and analyzed in the third section. In particular we study two
variants of the model and demonstrate that only shear force correlated vessel
collapse inside the tumor yields percolating morphologies of the  
tumor vasculature that is in good agreement with the experimental
observations. Further, a fractal analysis of the resulting tumor vasculature 
is performed to quantify the network architecture. Section~\ref{summary} summarizes our findings.

\section{Definition of the model}
\label{def_model}
The model introduced in \cite{Bartha06} and refined in
\cite{Lee06,Welter07} consists of a network of pipes
representing the vasculature and a cellular automaton, reminiscent of
the Eden model, representing the tumor.  Both parts interact via a
oxygen and/or nutrient concentration field, whose sources are perfused
blood vessels, and a growth factor concentration field, whose sources
are tumor cells. Both concentration fields are governed by diffusion
equations. Blood flow through the blood vessels is modeled as ideal
pipe flow with flow conservation in each node (junction) of the
network. New vessels are inserted into the network under certain
conditions, among which local growth factor concentration is the most
important one. Vessels can vanish (regress) once they are overgrown by
tumor cells. The latter proliferate when the oxygen/nutrient
concentration is sufficiently large and die when it is too low. This
model displays, over a wide range of the parameters, a growing tumor
whose radius increases linearly with time. It displays a
compartmentalization into a highly vascularized perimeter, a periphery
region with increased vessel density and a necrotic core, i.e.\ a 
center in which most of the tumor cells are dead and which is threaded only by a few
thick vessels surrounded by cuffs of tumor cells.

\begin{table}
  \caption{List of variables}
  \label{table1}
  \begin{tabular}{|cl|}
    \hline\noalign{\smallskip}
    Variables & Description \\
    \noalign{\smallskip}\hline\noalign{\smallskip}
    $L$ & System size\\
    $l(e)$ & Vessel length \\
    $t_\mathrm{c}$ & Critical life time of vessel\\
    $R_\mathrm{T}$ & Tumor radius \\
    $p$ & Vessel collapse probability\\
    $P(e)$ & Blood pressure at a given point\\
    $Q(e)$ & Blood flow per unit time\\
    $F(e)$ & Shear force on the vessel wall\\
    $MVD$ & Micro-Vascular Density\\
    \noalign{\smallskip}\hline
  \end{tabular}
\end{table}

\begin{table}
  \caption{List of parameters (l.u.=lattice unit, a.u.=arbitrary unit)}
  \label{table2}
  \begin{tabular}{|cll|}
    \hline\noalign{\smallskip}
    Parameters & Description & Values\\
    \noalign{\smallskip}\hline\noalign{\smallskip}
    $a$ & Initial vessel length & 8 (l.u.)\\
    $d(e),d_\mathrm{max}$ & Vessel radius, Maximum value & 1, 3.5(a.u.)\\
    $R_\mathrm{T}^0$ & Initial tumor radius & 30 (l.u.)\\
    $\Delta_\mathrm{angio}$ & Width of angiogenesis & 15 (l.u.)\\
    $p1,p2$ & Angiogenic probabilities & 0.01-0.5, p1/10\\
    $F_c$ & Critical shear for vessel collapse & 0.5-0.7\\
     \noalign{\smallskip}\hline
  \end{tabular}
\end{table}

In the present study, we simplify the model drastically by discarding the
oxygen and growth factor concentration fields. Unlike in~\cite{Bartha06,Lee06},
where the tumor was represented by an eden cluster, we consider a circular
tumor and linearly 
increasing radius with time. A highly vascularized region is observed in the
proximity of the tumor's periphery~\cite{Bartha06,Lee06} which arises due to the
interplay between the tumor secreted growth factor and existing vessel
network. Inside the tumor core, vessels die as a result of complicated
interactions with the tumor secreted growth factors and hypoxia~\cite{Hainaud06,Erber06}. Our
simplistic model, without considering such microscopic details, mimics the
essence of prior models~\cite{Bartha06,Lee06} by compartmentalizing the entire tumor
into an annulus at the tumor periphery where new vessels can emerge and an
inner annulus, where vessels can vanish with specialized criteria as discussed
below. Also the processes by which new vessels are inserted is simplified to
the extent of stochastic vessel growth. The model is defined as follows:

The model variables and model parameters are summarized in Tables~\ref{table1}
and \ref{table2}, respectively. The system configuration (tumor and network)
is defined on a 2-dimensional graph $G=(V,E)$, where $V=\{v\}$ denotes the set of
nodes and $E=\{e\}$ refers to the set of edges. Edge $e$, composed of ECs,
describe vessels and node is marked by the junction where two or more vessels 
intersect. For simplicity we take the edges to be parallel to the
coordinate axes. If $\vec{r_1}(x_1,y_1)$ and $\vec{r_2}(x_2,y_2)$ be the
position vectors of the  two end points along a vessel, then its length (edge
length) is given by, $l(e) := |\vec{r_1}(x_1,y_1) -\vec{r_2}(x_2,y_2) |$, with
$x_1=x_2$ or $y_1=y_2$.  The graph $G$ is then mapped onto an $L\times L$ (= $N$ sites)
square lattice. The lattice spacing $a$ denotes the initial vessel length which
corresponds to $(L/a)^2$ unit square-plaquettes (unit boxes) in the entire lattice.   

Since our main focus is on the vasculature in the neighborhood
of a growing tumor, and not on the structure of the tumor itself, the latter is only
virtually present in the system. But its effective interaction with the
vasculature is taken into account. The tumor $T$ is essentially represented by the
circular region of radius $R_\mathrm{T}$ centered at the lattice center.

In our model, blood vessel segments are represented by the edges. Each vessel
is compared with a cylindrical rod of uniform circular 
cross-section of radius $d(e)$. The blood flow through the vessel is
approximated by the laminar steady Poiseulles flow of a homogeneous,
incompressible fluid (Newtonian fluid). Hence the amount of fluid
$Q(e)$ that can pass through the vessel of length $l(e)$ per unit time is given by,  
\begin{equation}
Q(e) = \frac{\pi}{8 \eta} \frac{d^4(e) \Delta P(e)}{l(e)} \propto  \frac{d^4(e) \Delta P(e)}{l(e)},
\label{eq_flow}
\end{equation}
where $\Delta P(e)$ is the pressure difference
between the two ends of the vessel and $\eta$ is the coefficient of dynamic
fluid viscosity. Since $\eta$ is assumed to be a constant, we renormalize the
flow by the factor $\pi / 8 \eta$. The normalized shear force $F(e)$
acting upon the vessel wall, is given by,
\begin{equation}
F(e) = \frac{d(e) \Delta P(e)}{l(e)}.
\label{eq_shear}
\end{equation}
The boundary condition for the pressure is chosen to establish a
homogeneous flow and shear force distribution through each vessel-segments of the 
original network. We assume, the top-left corner ($x=1, y=1$) of the lattice at
pressure $P_0 $ (=1) is connected to the source artery, and the bottom-right corner
($x=L, y=L $) at pressure 0 is connected to the sink vein. The pressure at any
point on the boundary (e.g., top and bottom boundaries, $x=1,..,L$ for
$y=1,L$; and left and right boundaries,  $y=1,..,L$ for $x=1,L$ is
given by: $P(x,y) = P_0[1 - (x+y)/(2L)]$.  Using Kirchhoff's Current Law,
given by the conservation of total current (compared with the blood flow in
the present case) at a network junction, the blood pressure at every node is
calculated. Substituting them in Eqs.~\ref{eq_flow}-\ref{eq_shear} one 
immediately solves for the flow and shear force at every point of the network. The
boundary condition we have chosen, ensures a global net flow in the diagonal
((1,1) to ($L,L$)) direction~\cite{Bartha06}. In the following, we will denote
the flow and the shear force of the normal (undistorted) vessel network by
$Q_0$ and $F_0$ respectively.

\subsection{Initial configuration}
The initial configuration of the system is described by uniformly
spaced vessels of unit radius ($d_0(e)=1$) in a square lattice
\cite{Bartha06} and a circular tumor-zone in the center. Initial Micro
Vascular Density (MVD) is measured in terms of inter vessel distance $a$: MVD =
MVD$_0 = \{e|\vec{r}(x_0-x,y_0-y),~~x,y \in (-a/2,a/2]\}$. Numerically, this
  is computed by estimating the total vessel length $l(e)$ (made by endothelial
  cells $e$), within an open box of width $a$. In our 
simulation we consider initial vessel length $l(e)=a=8$ and hence $MVD_0 =
16$, which is half of the perimeter of a unit square plaquette of side
$a$. Since, initially $d(e)=1$ for the entire vasculature, as per
Eqs.~\ref{eq_flow}-\ref{eq_shear}, both $Q_0$ and $F_0$ take values $\Delta
P(e)/l(e)=0.5/L$. The circular tumor, centered at ($L/2, L/2$), has the initial
radius $R_\mathrm{T}^0 = L/20$. With these initial configurations, we now
proceed to describe the following deterministic and stochastic rules which
update the tumor and vasculature in each time step.


\subsection{Tumor Growth}
At each time step, the radius of the tumor $R_T$ is increased by one lattice site,
i.e., all sites out side the surface of the tumor with
$\vec{r}=\{\vec{r}|\vec{r}\notin EC, TC; \vec{r'} \in TC; |\vec{r} -
\vec{r'}| = R_\mathrm{T}(t+1)-R_\mathrm{T}(t)\}$ are occupied by tumor cells
at time $t+1$. Although, the proliferation of TC must be supported by availability of adequate
oxygen in the neighborhood of the tumor surface
\cite{Bartha06,Lee06}, in our simplified model, such restrictions are not
imposed. We assume that there is always sufficient
oxygen available in the neighborhood and along the tumor perimeter region to
assist the proliferation process. Tumor continues to evolve until the finite
size of the simulating system restrict its growth at
$R_\mathrm{T}=L/2-\Delta_{\mathrm{angio}}$, where $\Delta_{\mathrm{angio}}$ is
the width of angiogenesis outside the tumor surface, as illustrated below. 

\subsection{Angiogenesis}
\begin{figure*}
\resizebox{2.0\columnwidth}{!}{%
 \includegraphics{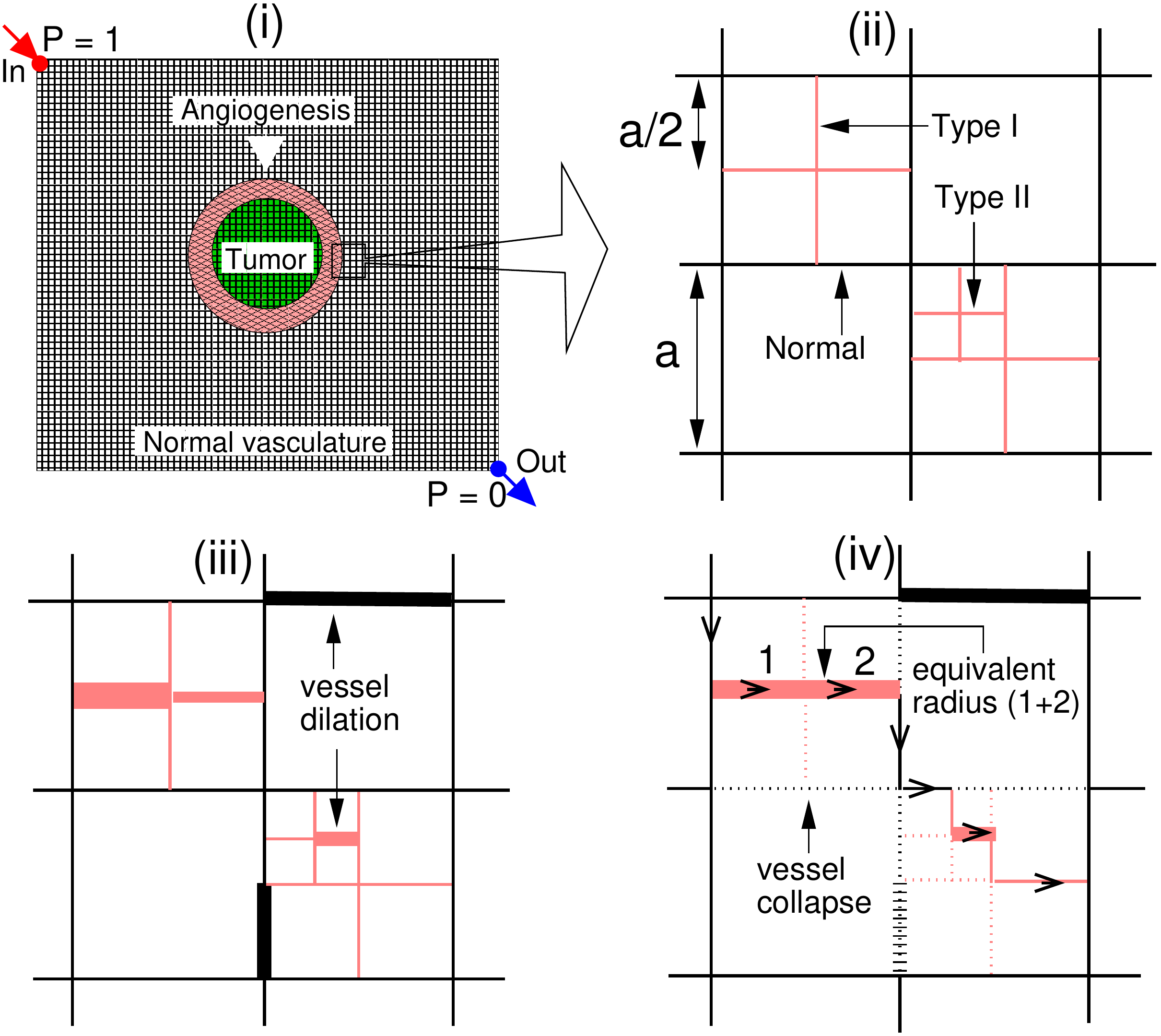}
 
}

\caption{(Color online) (i) Schematics of a tumor, angiogenic and normal vasculature. In our
 model the blood flow, if any, enters the system at the top left corner and
 exit from the bottom right corner. (ii) Magnified versions of the normal
 vessels and angiogenesis by inserting ``+'' shaped plaquettes within the
 unit (edge length $a$) and subunit cells (edge length $a/2$). (iii) Vessel
 dilation, shown by thick lines. (iv) Vessel collapse (dotted lines), i.e.,
 disappearance of vessel existing in (ii) and (iii). Vessels marked with 1 and 2 of
 different radii may be replaced by a single vessel with an equivalent uniform
 radius. After a number of vessel collapse, a single, long, non-uniform vessel might appear
 (shown by successive arrows).} 
\label{angio_dilation_collapse}
\end{figure*}

Each tumor cell releases some growth factor (GF) which activates the
proliferation of new blood vessels~\cite{Carmeliet00}. GF concentration
is high inside the tumor and decays to a negligible concentration beyond a distance
$R_\mathrm{T}+ \Delta_{\mathrm{angio}}$. Naturally, $\Delta_{\mathrm{angio}}$
is the range of angiogenesis. Since, the space inside the tumor is densely
occupied by tumor cells, the vessel proliferation in that region is highly
suppressed and therefore we consider the angiogenesis 
to occur mainly within the arbitrarily chosen annular ring of
thickness $\Delta_{\mathrm{angio}}=R_\mathrm{T}^0/2$, outside the tumor
surface. (We have repeated our simulation by considering the vessel
proliferation within the annular ring or thickness
$R_\mathrm{T}-\Delta_{\mathrm{angio}}/2$ to
$R_\mathrm{T}+\Delta_{\mathrm{angio}}/2$ and found that the choice of
$\Delta_{\mathrm{angio}}$ does not affect the tumor morphology in large time
limit: see~\ref{appD}). New vessels are added into the system in the form of a
``+'' shaped plaquette. A random unit cell (box of edge length $a$) is chosen
within the angiogenic regime, such that it is not occupied by an endothelial cell
($e=0$) and then a ``+'' plaquette (type I) is embedded into unit the cell with
probability $p_1$ (see Fig.~\ref{angio_dilation_collapse}). The insertion of
type I plaquette splits a unit cell into four equal sub-cells
of edge length $a/2$. At each time
step this process is repeated $\pi [R_{\mathrm{angio}}^2-{R_\mathrm{T}}^2]/a^2$
times which is equal to the total number of places where angiogenesis can
possibly occur. Once a type I plaquette is embedded into a unit cell, it is divided 
into four equal sub-cells each with edge length $a/2$. The substructure is made
more microscopic by inserting a second kind of plaquette, type
II, of edge length $a/4$ with probability $p_2$. Note that, a type II plaquette
can be placed into the system, only when a type I plaquettes is available in
the system. In this paper we present results for $p_1=0.05$ and $p_2=0.005$,
unless otherwise specified. For these parameter values the resulting pattern
of the vasculature appears to be similar to the one observed in physiological
circumstances. A comparative study for different angiogenic probabilities have
been discussed in section~\ref{fractalrobustness}. In this study, we do not
consider dynamic variations in angiogenesis during the course of tumor
evolution, i.e. both  $p_1$ and  $p_2$ are constant in time. This 
approximation is reasonable, but might not be the case in real systems and
should be addressed in our future models. 

\subsection{Vessel Dilation and Collapse}
Inside the tumor angiogenic sprouting is minimized, however, endothelial cell
proliferation still occurs along the vessel perimeter, leading to their enhanced
circumferential growth. Experiments proposed that high growth factor
and hypoxia induced activation of Eph/ephrin pathways~\cite{Hainaud06,Erber06}
might be responsible for vessel thickening inside tumor core. Thus in the
present model, all vessels inside the tumor are updated
sequentially by increasing 
their radius $d(e)$ at each time step by $\delta_d$ 
with a probability $p_d$ as long as $d(e)$ does not saturates to a maximum
value $d_\mathrm{max}$: $d(e) := d(e) + \delta_d \Theta(\chi - p_d)$, if
$d(e)<d_\mathrm{max}$, where $\chi$ is a random number and
$\Theta(x)=1$ for positive $x$ and 0 elsewhere.

Tumor vessels are tortuous and leaky. Vessel membranes become unstable due the
lack of pericytes, resulting in frequent vessel collapse and regression
\cite{Bartha06}. A reduced flow may not be sufficient to
prevent the vessel against the stress exerted by the neighboring tumor cells
and can become the key factor for their collapse~\cite{Goedde01}. We calculate
the vessel shear force and 
identify weakly perfusing vessels inside the tumor: $e \in E =
\{|\vec{r}-\vec{r'}|< R_\mathrm{T}\}$. Vessels with normalized shear forces
$F(e)/F_0$ falling below the critical value $F_\mathrm{c}$, are
removed from the rest of the vasculature with probability
$p$~\cite{Bartha06,Lee06}. Instead of removing the vessels by shear force
criterion, if we remove them with normalized flow
$Q(e)/Q_0<Q_\mathrm{c}$(not shown), or  normalized pressure (see
Appendix~\ref{appC}) a completely different vasculature is
observed. Other than the flow, pressure and shear force correlated vessel
collapse we have also employed another criterion for vessel collapse, in which
the vessels inside the tumor, below a certain age ($t_\mathrm{c}$=critical
lifetime), are removed probabilistically. This mechanism suggests that, if
vessels are able to survive until a critical life time $t_\mathrm{c}$, they
live for ever.

Both, vessel-dilation and vessel-collapse together give rise to long
nonuniform vessels as shown in
Fig.~\ref{angio_dilation_collapse}(iv). To make the calculations easier,
non-uniform vessels with different radii $d_i(e)$ (see vessel 1 and 2
in Fig.~\ref{angio_dilation_collapse} (iii)-(iv)) are assumed to be replaced by a
single vessel with effective radius $d_\mathrm{eff}(e) = [\sum_i
1/d_i^4]^{-1/4}$.

\section{Numerical Results}
The update rules described in the previous section, define a stochastic
process which we study numerically. At time $t=0$, we start with a
system of regularized vascular network and a pre-existing circular tumor at the
center. We have simulated our model over a wide range of parameters
and applied various criteria for vessel collapse. In the
following, however, we present results which produce realistic 
vessel morphologies reported in prior investigations~\cite{Doeme02}.

\subsection{Random vessel collapse inside the growing tumor leads to a ``void'' core} 
First we consider the growth of a circular tumor in the back-bone of a
capillary network arranged in a 2$d$ square lattice, as described earlier. The entire
lattice is assumed to be connected to outer vessels (vein and arteries) along
one of the two diagonal directions. During the tumor evolution, vessels inside
the tumor collapse with a certain probability, resulting in many non-circulating
dangling vessels. At each time step these non-circulating
(non-biconnected) vessels are completely removed from the rest of the
lattice. Vessels do not collapse further if they survive up to a critical
lifetime $t_\mathrm{c}$. It is easy to visualize that in the beginning,
although all intra-tumoral vessels are in critical state, with the maturation
(aging) of the tumor, critical zone moves away from the tumor center and stay
limited within the annulus of thickness $(t-t_\mathrm{c})$ (for
$t>t_\mathrm{c}$) under the tumor surface. When the tumor diameter becomes comparable to the
linear size $L$ of the lattice, we examine if a continuous path exists inside
the tumor along any of the two diagonal directions of the lattice. If such a path
exists, the probability $P_{\infty}$ of finding a spanning cluster is 1, and
0, otherwise. Slowly increasing the vessel collapse probability $p$ from zero
to a maximum value, we calculate $P_{\infty}$. A transition of the vasculature
morphology from a dense to a void network is observed at the critical
$p=p_\mathrm{c}$. Two different cases have been studied in this context:
vasculature I. {\it without} and II. {\it with} angiogenesis.  

\begin{figure*}
  \resizebox{2.0\columnwidth}{!}{
    \includegraphics{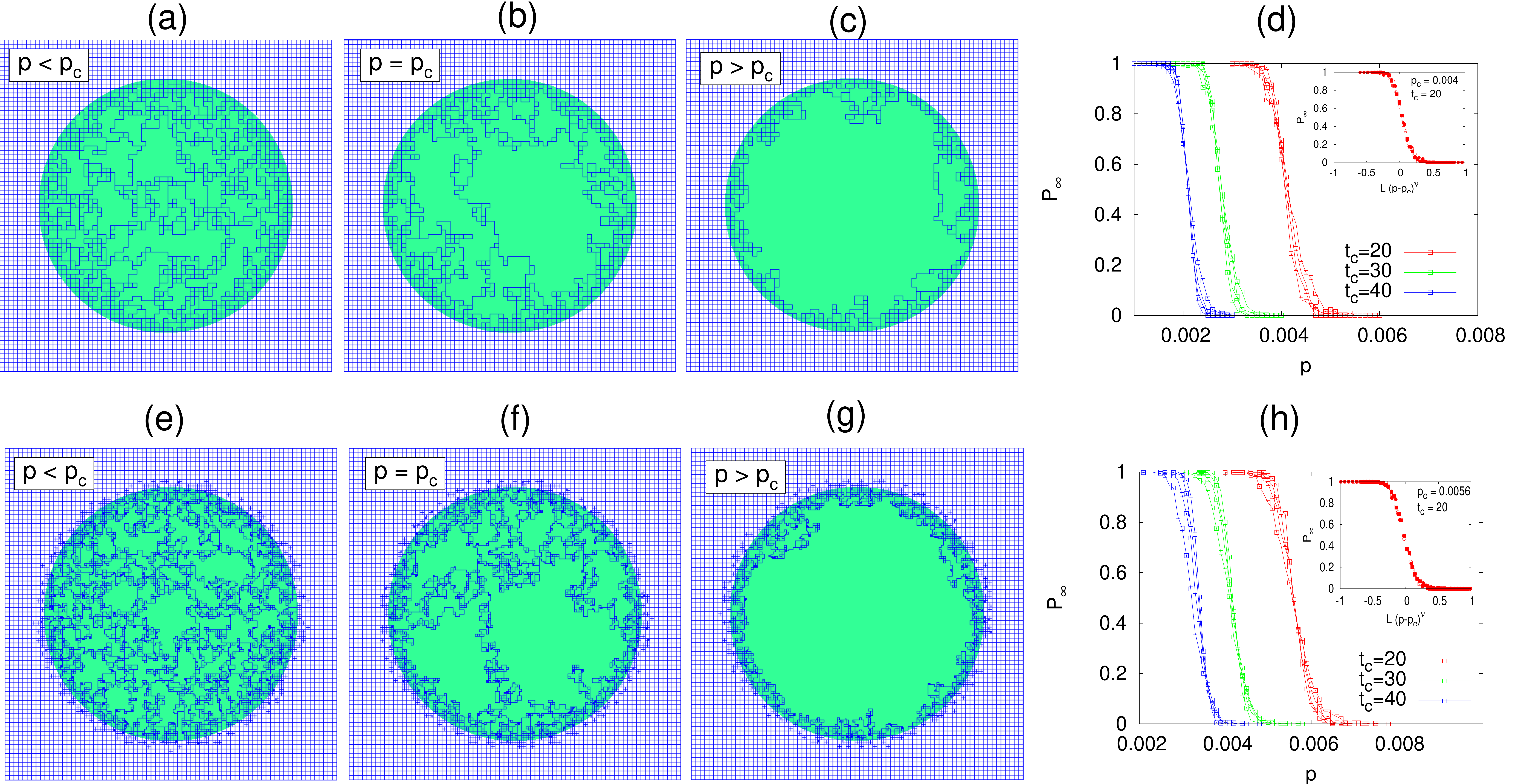}
  }
  
  \caption{(Color online) {\bf Top row:} Snapshots of a growing tumor vasculature for random
    vessel collapse without angiogenesis. Circular green region represents the
    tumor and blue lines represent the blood vessels. Morphologies correspond to different
    collapse probabilities (a) below (b) at and (c) above the percolation 
    threshold. In the subplot (d) $P_{\infty}$ is shown as a
    function of collapse probability $p$ for different values of critical
    vessel lifetime $t_\mathrm{c}=$20, 30 and 40 with systems
    sizes $L=$320, 400, 480 and 600. As $t_\mathrm{c}$ increases, vessel stabilization
    takes longer and the percolation threshold $p_\mathrm{c}$
    goes down. {\bf Inset (d):} Data collapse of the curves
    for $t_\mathrm{c}=20$ with $p_\mathrm{c} \simeq 0.004$ 
    and $\nu=4/3$. {\bf Bottom row:}(a) Snapshots of a growing tumor vasculature with
    angiogenesis. Probabilities for plaquettes are $p_1=0.05$ and
    $p_2=0.005$. Subplots (e), (f) and (g) represents vasculature 
    below, at and above the percolation threshold respectively. In (h) $P_{\infty}$
    vs. $p$ are shown for the same $L$ and $t_\mathrm{c}$,
    mentioned above. {\bf Inset (h):} Scaled version of the data for
    $t_\mathrm{c}=$20 with $p_\mathrm{c} \simeq 0.0056$ and $\nu=4/3$.}  
  \label{Pinf_snpshot_without_angio}
\end{figure*}

{\bf{I. Vasculature without angiogenesis disappear at small collapse
    probability}:} Morphologies of the tumor vasculature for random vessel 
collapse and without angiogenesis are displayed in
Fig.~\ref{Pinf_snpshot_without_angio} (a), (b) and (c), for   
collapse probabilities below, above and at the percolation threshold
respectively.  
We calculate $P_{\infty}$ for different $L$, as a function of vessel collapse
probability $p$. The results are plotted in Fig.~\ref{Pinf_snpshot_without_angio} (d)
for $t_\mathrm{c}=20, 30, 40$. There is no restriction imposed on the choice
of $t_\mathrm{c}$, since we have also observed percolation for various other
values of $t_\mathrm{c}$. At small $t_\mathrm{c}$, quick vessel stabilization
leads to a diverging percolation thresholds, whereas, at large $t_\mathrm{c}$,
rapid vessel collapse results in a converging percolation threshold. Thus, for
the sake of clarity, we have not shown those data in Fig.~\ref{Pinf_snpshot_without_angio}. All our 
data, presented in Fig.~\ref{Pinf_snpshot_without_angio} are averaged over 400
ensembles. A sharp decrease in the value of $P_{\infty}$
characterizes the percolation transition at $p_\mathrm{c}$, above which a spanning
cluster ceases to exist in the system. For $t_\mathrm{c}$=20,
$p_\mathrm{c}=$0.004, which decreases further as $t_\mathrm{c}$ is increased. The
exponents $\nu=4/3$ has been estimated from the finite size scaling analysis
for all $t_\mathrm{c}$ (see inset of Fig.~\ref{Pinf_snpshot_without_angio} (d)
for $t_\mathrm{c}=20$) and its value is found to be independent of $t_\mathrm{c}$.

{\bf{II. Vasculature with angiogenesis remain percolating at higher collapse
    probability}:} Here we introduce angiogenenic
sprouting in the tumor surrounding vasculature by inserting type I and II
plaquettes. The resulting tumor is found 
to be enclosed by a shell of high MVD, as depicted in
Fig.~\ref{Pinf_snpshot_without_angio} (e), (f) and (g). In a similar way, like
in the previous case, we carry out measurement to find $P_{\infty}$ for
angiogenic probabilities $p_1=0.05$ and $p_2=0.005$ and plot the results in
Fig.~\ref{Pinf_snpshot_without_angio} (h). The value of  
$p_\mathrm{c} \simeq$ 0.0056 and $\nu=4/3$  are estimated from 
the data collapse for $t_\mathrm{c}=$ 20, as shown in the inset of
Fig.~\ref{Pinf_snpshot_without_angio}(h). It is readily observed that
percolation transition for $t_c=20$, in presence of angiogenesis, is shifted
to a higher value at $p_\mathrm{c}$=0.0056, from $p_\mathrm{c}$=0.004, when
there is no angiogenesis.   

Our results corresponding to Fig.~\ref{Pinf_snpshot_without_angio} suggest
  that random vessel collapse with and without angiogenesis yields two
  distinct phases of the tumor vasculature: one in which the tumor center is completely
vascularized and one in which it is completely void,
separated by a percolation transition at a critical collapse
probability. Both phases appears to be unrealistic compared to the real tumors
and prior research~\cite{Bartha06,Lee06}, except
the critical point itself when a few vessels threading the tumor center. This
vasculature is similar to the reminiscent of a real tumor vasculature - but requires, in this
model, a fine tuning of a certain parameter (the collapse probability),
which is again unrealistic. In the following, we will present a mechanism by
which the vessel network drives itself into such a critical state.

\subsection{Shear force correlated vessel collapse inside the growing tumor
  leads to a ``percolating'' vasculature} 
\begin{figure*}
  \resizebox{2.0\columnwidth}{!}{%
    \includegraphics{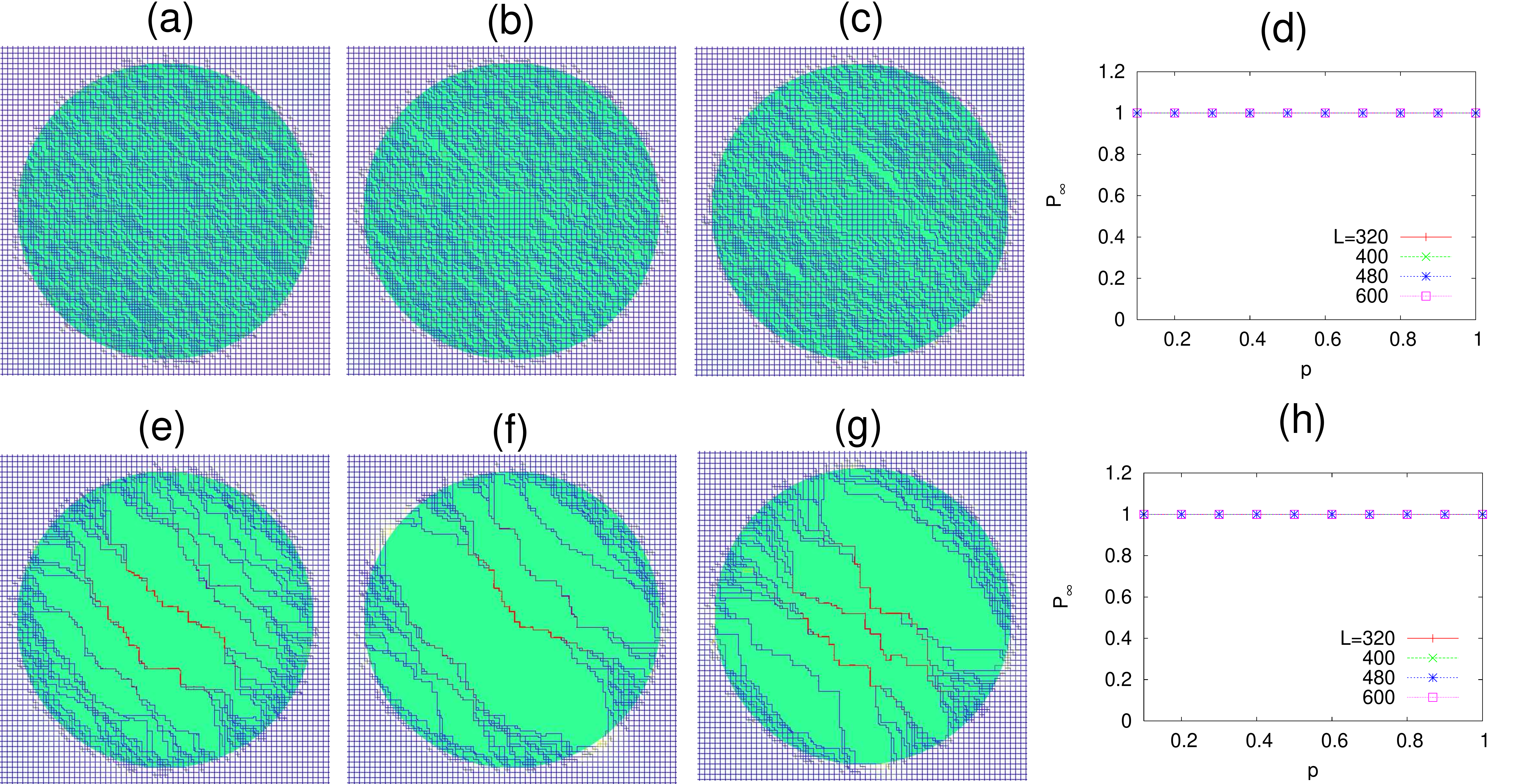}
  }
  \caption{(Color online) Snapshots of a growing tumor
    vasculature for shear force correlated vessel
    collapse. Tumor is shown in green. Blood vessels are colored
    according to their instantaneous blood flow: red indicates
    high (inside the tumor, along the flow direction: top-left
    to bottom-right), blue indicates normal (away from the
    tumor surface) and yellow indicates low blood flow. {\bf Top row:} Uniform
    vessel inside tumor. Morphologies 
    correspond to different collapse probabilities: (a) small
    (b) medium and (c) large. In the subplot (d) $P_{\infty}$
    is shown as a function of collapse probability $p$ for
    $F_c=0.5$ with system sizes $L=$320, 400, 480 and
    600. {\bf Bottom row:} Dilated vessel inside 
    tumor, similar to the Top row. Vasculature is always
    percolating inside the tumor.} 
  \label{Pinf_snpshot_vesDia}
\end{figure*}

We carried out a similar study of percolation, as described in the previous section,
for the shear force correlated vessel collapse. Here we consider two different cases: tumor
vasculature I) with uniform vessel and II) dilated vessel. In these cases, vessels neither undergo random collapse, nor stabilize
automatically with their age ($t_c$), as considered in the previous
section. (Note that, aging increases the stability of the vessels. To verify
the robustness of the vasculature in presence of aging, simulations are carried
out and the results are shown in Appendix~\ref{appE}.) Vessels are removed
with normalized shear force falling below the critical value $F_c=0.5$, and
with collapse probabilities $p$ varying from 0 to 1. The results are depicted
in Fig.~\ref{Pinf_snpshot_vesDia}. In 
contrast with the previous results (void tumor at high $p$, as depicted in
Fig.~\ref{Pinf_snpshot_without_angio}), our present analysis asserts that tumor
interior is always percolating, no matter what the magnitude of the collapse
probabilities are.   

We now focus on the microstructures of the angiogenic vasculature evolving
under shear correlated vessel collapse.   

\begin{figure*}
  \resizebox{2.0\columnwidth}{!}{%
    \includegraphics{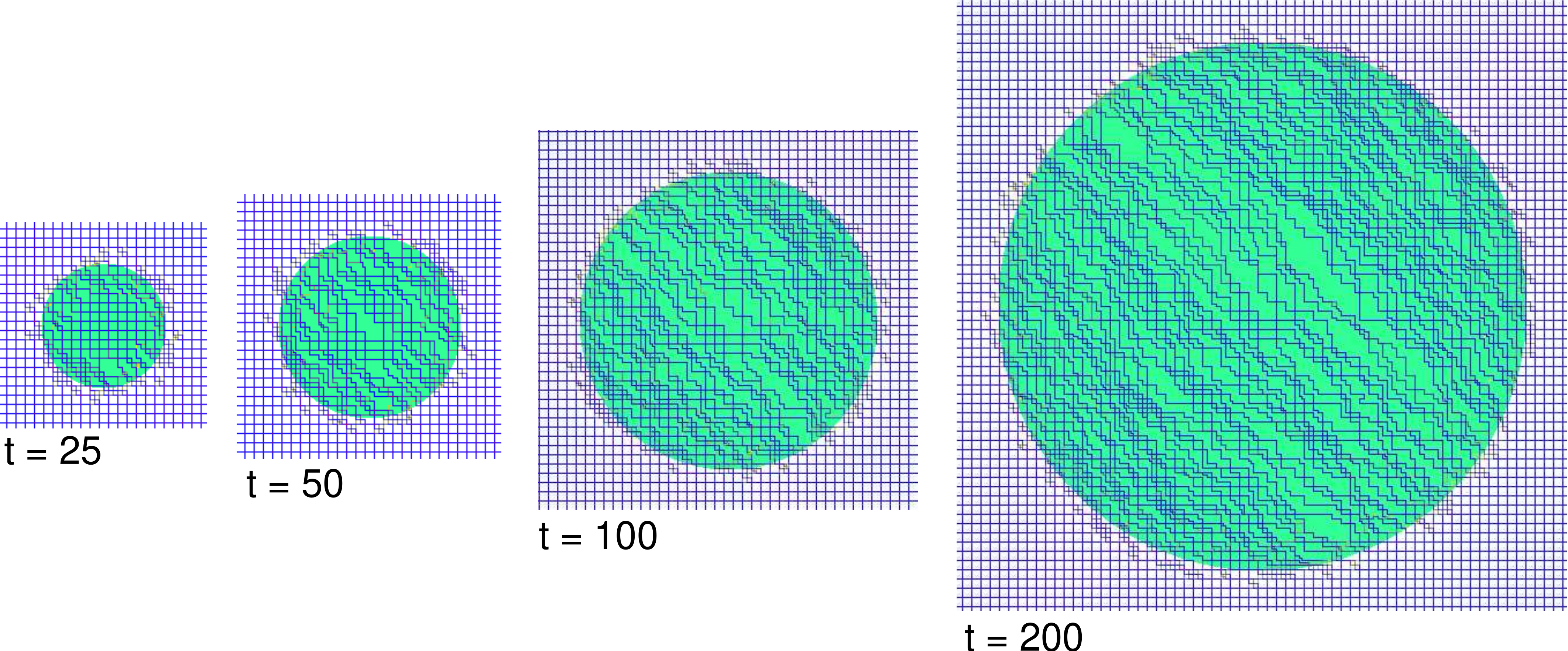}
  }
  \caption{(Color online) Snapshots of a growing tumor with uniform vessel
    radius and angiogenic probabilities $p_1=0.05$ and $p_2=0.005$ for linear
    system size $L=600$. Vessels are collapsed under critical shear
    $F_\mathrm{c}=0.7$ with probability 0.025. The color code is described in
    Fig.~\ref{Pinf_snpshot_vesDia}.}  
\label{evol_uniform_vessel}
\end{figure*}

{\bf{I. Lack of vessel dilation produces dense tumor vasculature:}}
First we discuss the case when there is no vessel dilation occurring inside
the tumor, i.e., vessel radius $d=1$ is a constant. Vessels inside the tumor with shear
force (as per Eq.~\ref{eq_flow}, $Q(e)=F(e)$, since $d=1$) falling under the critical
value $F_\mathrm{c}=F(e)/F_0=0.5$ are removed with probability
$p=0.025$. In Fig.~\ref{evol_uniform_vessel} we show the
snapshots of the evolving vessel network. 

As time progresses, homogeneous vascular network with constant MVD throughout,
is rapidly changed into an inhomogeneous one (see
Appendix~\ref{appA}). Angiogenesis in the peritumoral 
region produces a  huge number of vessel junctions, each of which divides the original flow
into multiple components. The new vessels carry less blood resulting in a weaker shear
force. Therefore they collapse as time goes by. Once the weak vessels are
removed, remaining vessels carry more blood and stabilize against the
collapse. In the late stage of the simulation, vessel strips exists along and
transverse to the principal flow direction, accompanied by empty regions. However, under
any circumstances, the final structure of the remaining vessels appear to be
dense and bear little resemblance to the actual tumor vasculature.

{\bf{II. Vessel dilation inside the tumor produces realistic tumor vasculature:}}
\begin{figure*}
  \resizebox{2.0\columnwidth}{!}{%
    \includegraphics{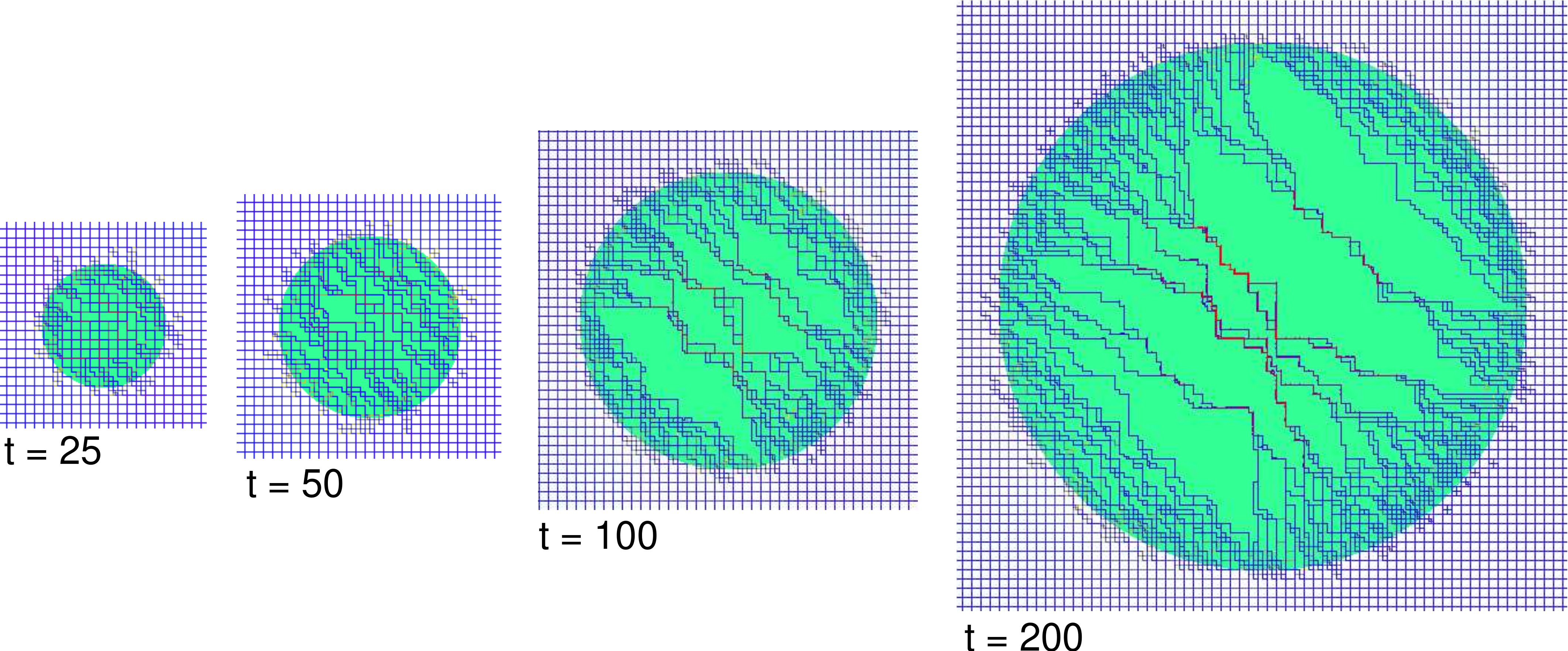}
}
\caption{(Color online) Snapshots of a growing tumor similar to
  Fig.~\ref{evol_uniform_vessel}, but with a probabilistic vessel dilation
 interior to the tumor. Vessels are collapsed under critical shear
 $F_\mathrm{c}=0.5$ with probability 0.025. The color code is the same as
 described in Fig.~\ref{Pinf_snpshot_vesDia}.}
\label{evol_nonuniform_vessel}
\end{figure*}
So far, we have presented results for tumor vasculature with uniform vessel
radius. The scenario becomes largely different once the vessels start
thickening inside the tumor simultaneously with the angiogenesis occurring out side the
tumor (see Appendix~\ref{appB}). Starting with a homogeneous network of vessels, 
with uniform radius $d(e)=1 \forall e$, probabilistic vessel dilation (with
probability 0.025) for vessels with $d(e)<d_\mathrm{max}=3.5$ are allowed to
occur, inside the tumor. Intra tumor vessels, with shear force (as per
Eq.~\ref{eq_shear}) falling under critical value $F_\mathrm{c}=F(e)/F_0=0.5$ are
removed with collapse probability $p=0.025$. Resulting vasculature morphologies are
shown in Fig.~\ref{evol_nonuniform_vessel} (for measurements MVD, shear force,
vessel radius, pressure gradient, see Appendix~\ref{appB}). One observes that, the interplay
between angiogenesis and shear correlated vessel collapse  
rapidly changes the homogeneous vascular network into an inhomogeneous
one. Resultant vessels inside the tumor grow thick and become stable against
any further collapse. 

 
\subsection{Flow correlated percolation influences the fractal vasculature}
\label{fractalrobustness}
\begin{figure}
  \resizebox{1.0\columnwidth}{!}{%
    \includegraphics{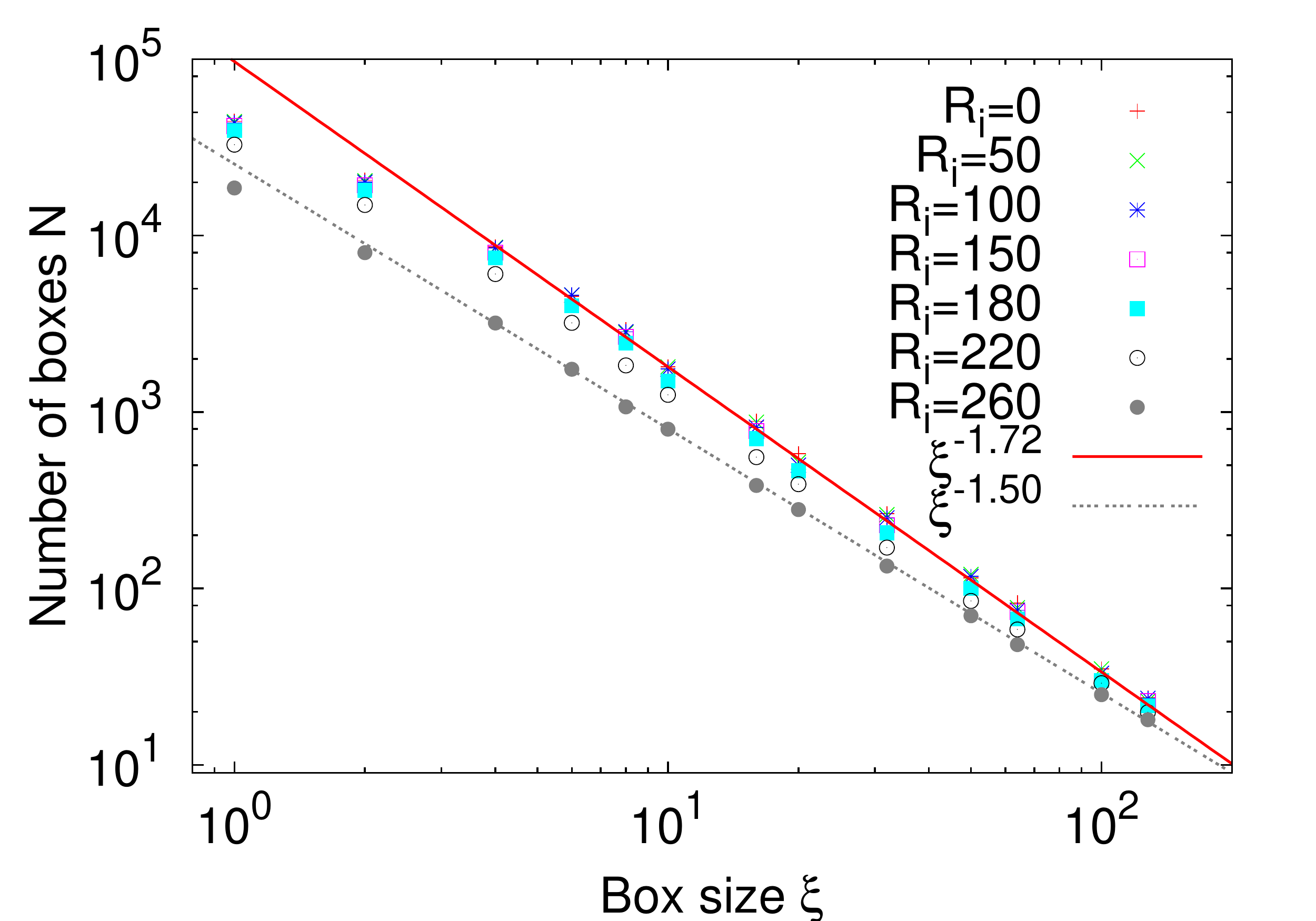}
}
 \caption{(Color online) Box counting estimate of $D_\mathrm{f}$ of the tumor
   vasculature. We confined the
   measurement to annuli with fixed outer radius that is determined by the
   limit of the peritumoral plexus (extend up to the angiogenic region outside
   the tumor periphery) and with varying inner radius $R_i$. The number of
   boxes of size $\xi$ needed to cover the vasculature within the annuli, is
   plotted in log-log scale. The slope of the curves correspond to $D_\mathrm{f}$
   which decreases with increasing $R_i: D_\mathrm{f}=1.72\pm0.01$ for
   $R_i=0$(corresponds to the complete tumor and angiogenic vasculature) and
   $D_\mathrm{f}=1.50\pm0.02$ for $R_i=260$ (corresponds to the angiogenic
   vasculature only), indicating that the fractal dimension is drastically
   modified inside the tumor.}  
\label{fig_fractdim}
\end{figure}

\begin{figure*}
  \resizebox{2.0\columnwidth}{!}{%
    \includegraphics{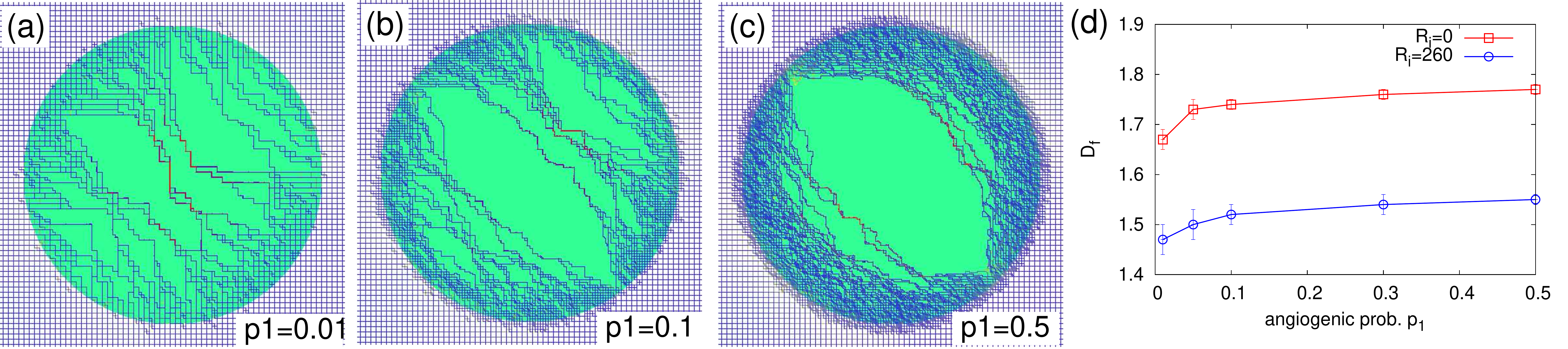}
}
 \caption{(Color online) Subplots (a),(b) and (c) shows tumor vasculature at
   time $t=220$ for different angiogenic probabilities $p_1$ marked inside the
   figures. In (d) fractal dimension $D_\mathrm{f}$ is plotted as function of
   $p_1$. The color code is the same as described in
   Fig.~\ref{Pinf_snpshot_vesDia}.}  
\label{fig_vasculature_fd}
\end{figure*}

This section focuses on the geometrical aspects of the tumor vasculature. In the
asymptotic time limit, one observes the turn over of a homogeneous vasculature
into a completely inhomogeneous one, resulting from the inter-play
between angiogenesis and shear force determined vessel collapse. One remains
with irregular vasculature with high MVD in the peritumoral neighborhood and a
few long, thick vessels in the interior of the tumor, 
prevailing along the flow direction. It is still unclear, what are the main 
influencing factors that reshape the tumor vasculature. A detailed insight
into the remodeling process of the vessel network would be given its fractal
dimension. A fractal, in general, is a rough geometrical structure which is
self similar or scale invariant. However, the emerging network in our study,
displays a broken symmetry due to the diagonal global flow and thus spatially
inequivalent regions arise. Estimation of the fractal dimension would shed
light on the magnitude of spatial influence applied by the tumor on the vessel
growth process. Therefore, it is worth to focus on the fractal 
properties and quantify the structural changes of the network in the 
tumor environment. The fractal dimension $D_\mathrm{f}$ of the vascular network is determined by the
box counting method~\cite{Mandelbrot83}. It is given by the ratio between the
logarithm of the number of boxes needed to cover the vessel network
($N_{\xi}$) and the logarithm of the linear box size:
\begin{equation}
D_\mathrm{f} = -\lim_{{\xi} \rightarrow 0} \frac{\ln \mbox{N}_{{\xi}}} {\ln {\xi}}.
\end{equation}
We analyze the modified vasculature extending from the center of the tumor up to the
angiogenic peritumoral regions. Beyond this region the vasculature is
normal. In our measurement, we divide our region of interest into several annuli with
fixed outer radius, determined by the limit of the peritumoral plexus, and
varying inner radius $R_i$. Fig.~\ref{fig_fractdim} 
displays the number of boxes $N_\mathrm{\xi}$ as a function of the box size $\xi$ in a
log-log scale. (Note that, the slope in
Fig.~\ref{fig_fractdim} decreases as the box size $\xi$ goes below the 
characteristic length scale of the system, set by the initial vessel length
$a=8$. Therefore, in the measurement of $D_\mathrm{f}$ those points are
disregarded.) Fitting a straight line through $\sim$ 11 data points (extending almost
two decades), we find $D_\mathrm{f} = 1.72 \pm 0.1$, corresponding to the
entire vasculature remodeled by the tumor. Similar studies, as a function of
the angiogenic probabilities result in little variation in $D_\mathrm{f}$,
although the morphology appears to be quite different (see
Fig.~\ref{fig_vasculature_fd}). Our numerical estimate of 
$D_\mathrm{f}$ could be compared with the experimental
prediction~\cite{Gazit98} obtained on a 2$d$ slice of human carcinoma using 2$d$
image analysis. Instead of studying the entire vasculature, if we consider  
annular rings of different thickness, near the peritumoral plexus we obtain $D_\mathrm{f} =
1.50 \pm 0.02$. We observe only small changes of $D_\mathrm{f}$, when the
angiogenic probability $p_1$ is varied from 0.05 up to 0.5, and
the range of angiogenesis $\Delta_{\mathrm{angio}}$ is varied from $a$ up to
$3a$ (not shown). 

{\bf{Robustness of fractal vasculature under varying angiogenesis:}}
Tumor vasculature is observed to be modified by the rate of
angiogenesis. Therefore a quantitative study of the vessel network is carried
out by varying the amplitude of angiogenesis. A similar study
reported the variations of $D_\mathrm{f}$ as a function of critical shear
force~\cite{Welter07}. Our study, so far, addressed angiogenesis occurring with probabilities
$p_1=0.05$ and  $p_2=p_1/10$. Keeping this ratio fixed, we now vary $p_1$ from 0.01 up to
0.5. The vasculature resulting at very late stage of the simulation are shown in
Fig.~\ref{fig_vasculature_fd}(a)-(c). With increasing angiogenesis, average
blood shear force decreases in the interior of the tumor, giving rise 
to a fusiform, avascular region along the principal flow direction. However,
vasculature through the tumor still remains percolating. We further
analyze the fractal dimensions $D_\mathrm{f}$ of the obtained vessel networks,
for which the results are plotted in Fig.~\ref{fig_vasculature_fd}(d). The fractal
dimension for both the entire tumor vasculature ($R_i=$0) and the annular
angiogenic region ($R_i=$260) remain almost constant in the
regime of large $p_1>0.05$, however, decays for smaller $p_1$. Our analysis
confirms the robustness of the tumor vasculature under angiogenic
perturbations.  

\section{Summary}
\label{summary}
Tumor induced angiogenesis is a basic mechanism in cancer
development. We have presented a simple stochastic model which
initiates with the regular vasculature that consists of capillaries of
equal diameter arranged in a regular grid with a given MVD ensuring a
homogeneous distribution of flow and constant shear stress in all
constituting vessels. Once the tumor grows up to a certain size, the
vasculature gets modified into characteristically different sections. Since the vessel
segments need to be biconnected to the exterior network, a large part
of the tumor vasculature can be cut from the rest of the network by
only a few vessel collapse. The inhomogeneity in the network extends
from high MVD with thin vessel region in the peritumoral plexus up to a
sparsely populated and thick vessel environment in the interior of the
tumor~\cite{Carmeliet00,Doeme02}. The spatio-temporal characteristic curves
for MVD, blood flow, shear force, pressure gradient and vessel radius as
predicted in ~\cite{Bartha06,Lee06,Welter07}, are reproducible by our
simplistic approach (see Appendix B). 

Our model predicts that inhomogeneity in the network is caused by the
interplay between the excessive vascularization in the tumor periphery and
vessel collapse in the interior. We have implemented two different
mechanisms for the probabilistic rupture of vessels {\it viz}, random
collapse and shear stress correlated collapse. The latter is motivated
by the study on normal vessels which undergo a structural reduction
of internal vessel diameter due to a decreased wall shear stress
(brought about by changes of blood flow)~\cite{Pries95}. When the vessels are
removed randomly, the interior of the tumor is either containing a dense
vasculature or completely empty. This phenomenon
is well understood from the basic law of percolation
theory~\cite{Stauffer94} and suggest the existence of a spanning
vasculature for a particular collapse probability $p=p_\mathrm{c}$, known as
percolation threshold. Under the random vessel collapse criterion, a
percolating phase requires a fine tuning of $p$ which depends strongly on the
model parameters.

Next we considered a probabilistic vessel collapse, with shear force
falling below a critical value inside the tumor, while the vessel diameter is
kept constant. After the initial removal of weakly perfusing vessels, which
are usually perpendicular to the flow direction, long vessels remain parallel
to the principal flow 
direction. These vessels may further suffer from the insufficient shear 
force due to their increased length. However, an increased pressure drop
across the vessel or an enlargement in vessel diameter can amplify the
amount of blood flow and consequently the shear force. Vessels, for which the
shear compensation is not made by the increased flow or pressure drop, 
eventually collapse. Thus, in the long run, a few empty regions appearing
along the principal flow direction, with increased MVD regions visible along
the transverse flow direction, producing a physically unrealistic tumor
vasculature, unlike as reported in
~\cite{Carmeliet00,Doeme02,Bartha06,Lee06,Welter07,Welter09}.  

Our model predicts a percolating morphology of tumor vasculature which is
maintained by the correlation between the probabilistic vessel collapse and
the local shear stress exerted by the blood flowing through it. However, they key
ingredient to produce a realistic vasculature is still missing.   

Therefore, we included the intratumoral vessel dilation. High growth factor and
hypoxia induced activation of Eph/ephrin pathways has been reported to reduce the angiogenic
branching and enhance the vascular circumferential growth inside the
tumor~\cite{Hainaud06,Erber06}. Vessel dilation which occurs inside the 
tumor~\cite{Carmeliet00,Doeme02} and taken into account in prior 
models~\cite{Bartha06,Lee06,Welter07,Welter09}, plays crucial role in
determining the vasculature morphology. Implementing probabilistic 
vessels dilation inside the tumor together with shear stress correlated vessel
collapse, a physically realistic 
vasculature is established~\cite{Bartha06,Lee06}, which is robust under a large parameter
variations. The inherent mechanism for the stability of the 
remaining vessels after a collapse event had occurred, caused by the
diversion of the blood flow and consequential increase in shear stresses,
bringing them above the collapse threshold.

Another potential candidate for vessel dilution inside the tumor could
naturally be the critical blood pressure ($P_\mathrm{c}$), above which vessels are
torn  apart. We have tested this scenario by varying the critical threshold of
of $P_\mathrm{c}$ from 0.001 to 0.5. The snapshots, displayed in
Appendix~\ref{appC}(Fig.~\ref{press_vessel_collapse}), shows that 
the resulting tumor vasculature, below moderate $P_\mathrm{c}$, is
percolating, but its morphology appears physically unrealistic.  

In this work, we have not considered dynamic variations in angiogenesis during the
course of tumor evolution, i.e. both  $p_1$ and  $p_2$ remained constant in
time. This approximation is reasonable, but might not be the case in
real systems and should be addressed in our future models. 

Our model also predicts a fractal geometry of the tumor vasculature. The fractal dimension
$D_\mathrm{f}=1.72 \pm 0.1$ that we have estimated, is compared with
$D_\mathrm{f}=1.89 \pm 0.04$~\cite{Gazit98} obtained by two-dimensional images
analysis of vessel networks in human carcinoma. Previous theoretical estimates of
$D_\mathrm{f}$ in 2$d$ ~\cite{Bartha06,Welter07} and in 3$d$
~\cite{Lee06}, however differ slightly from our results. It has been argued
that extracellular matrix inhomogeneity in tumors might be responsible for the invasion
percolation~\cite{Furuberg88,Sheppard99} and fractal architecture of 
tumor vasculature. Our analysis and those performed earlier
\cite{Bartha06,Lee06,Welter07}, having no such extracellular matrix
dependency, suggests that the flow correlated percolation could also be the
possible origin of the fractal tumor vasculature. To confirm this,
extensive simulations should be carried out with bigger systems. Our present model does not assume
any correlation between length, radius and thickness of the vessel. Such 
correlation might be relevant in a real context and therefore should be addressed in our
future studies.

\begin{acknowledgements}
We wish to thank H. Rieger and K. Bartha for useful discussions.
\end{acknowledgements}

\begin{appendix}
\section{Shear correlated vessel collapse without vessel dilation: quantifying
  MVD, blood flow, shear force, pressure field}  
\label{appA}
The entire dynamical process is quantified by measuring the following
quantities at each time step: average MVD is measured as a function of radial
distance $r$ from the center of the tumor by estimating the total length of
endothelial cells $l(e)$ spanning through a unit cell (a unit box of edge
length $a$, as defined earlier) within the annular ring of radius $r$
and thickness $\delta r=4$. Radial distribution of flow and shear force are
also calculated in the similar way and pressure is calculated at each point
$(x,y)$ over the entire lattice.   

\begin{figure*}
  \resizebox{2.0\columnwidth}{!}{%
    \includegraphics{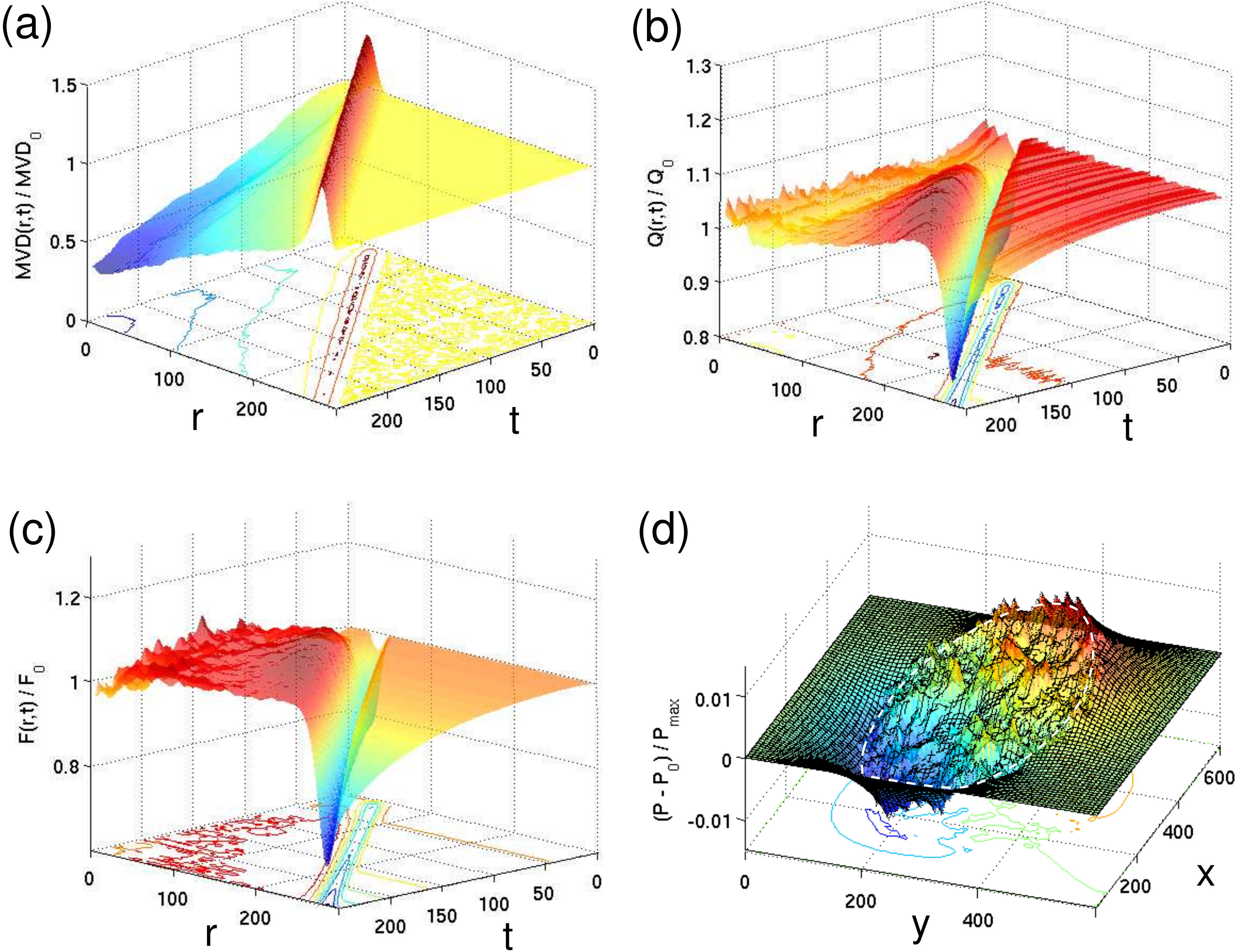}
}
\caption{(Color online) (a) Normalized micro vascular density MVD(r,t), defined as the average
 number of vessels per box of side $a$, with respect to the normal tissue
 MVD$_0$. The average is done over an annulus of width $a=8$
 with central radius $r$. Subplots show: (b) normalized blood flow per vessel and (c) 
 normalized shear force on the  vessel walls. In (d) we
 show the difference in the blood  pressure field $P(x,y)$ at time $t=150$
 from its normal value at time $t=0$. The tumor is enclosed by the dotted
 region. The global flow direction,  enforced via the boundary conditions, is
 from left (1,1) to right 
 (600,600). Looking along this direction, the pressure is decreased in front
 of the tumor(left) and increased through the tumor until the exit end (right).} 
\label{main_plot_uniform_vessel}
\end{figure*}

The change in normalized MVD (i.e., with respect to MVD$_0$) is plotted as a
function of radial distance $r$ and time $t$ in
Fig.~\ref{main_plot_uniform_vessel} (a). A sharp maximum is 
observed at a distance $r=R_\mathrm{MVD}(t)$ which is approximately equal to the tumor radius
$R_\mathrm{T}(t)$. Due to the fixed angiogenic probabilities $p_1$, $p_2$ and
no vessel regression out side the tumor periphery, maximum MVD remains
constant in time. Since the tumor is assumed to grow linearly in time, 
$r=R_\mathrm{MVD}(t)$ is also found to be a linear function of $t$. Far out
side the tumor ($r\gg R_\mathrm{MVD}$), the normalized 
MVD remains constant, due to the absence of angiogenesis. For radii
$r<R_\mathrm{MVD}(t)$ (inside the tumor), the MVD decreases very
quickly to the normal tissue MVD$_0$ and then slowly to values slightly
lower than MVD$_0$ towards the center of the tumor. 

The normalized blood flow $Q(r,t)/Q_0$ is presented in
Fig.~\ref{main_plot_uniform_vessel} (b). The profile shows a small dip 
around $R_\mathrm{MVD}(t)$, and then increases sharply to a value similar to the  
normal tissues and remains almost constant inside the tumor. The sharp fall
results from the effect of high MVD around 
the periphery  of the tumor which divides the flow into many components. On
the contrary, a constant normalized flow inside the tumor is caused by the
uniformity of vessel radius. Similar 
behavior, like the blood flow, has also been displayed by the normalized shear
force $F(r,t)/F_0$, as shown in Fig.~\ref{main_plot_uniform_vessel} (c). Later
we will see, in Appendix~\ref{appB}, that a variable radius changes the entire
scenario.  

Morphological changes in the vasculature modifies the pressure field in the
network. E.g., pressure field at time $t=150$ with respect to the one at
$t=0$, as reported in Fig.~\ref{main_plot_uniform_vessel} (d), shows no
difference away from the tumor region, however, it decreases near the periphery of
the tumor. This characteristic feature arises as the total blood flow enters
the tumor through the highly vascularized peritumoral region, resulting in a
decreased pressure gradient. To preserve the flow conservation, the pressure
gradient increases inside the tumor till the exit end. If the critical shear force is tuned
to a higher value ($F_\mathrm{c}>0.5$), the pressure difference flips it's
pattern (not shown) due the lack of MVD on both sides of the tumor along the
flow direction.

\section{Shear correlated vessel collapse with vessel dilation: quantifying
  MVD, blood flow, shear force, pressure field, average vessel radius}   
\label{appB}
\begin{figure*}[h]
  \resizebox{2.0\columnwidth}{!}{%
    \includegraphics{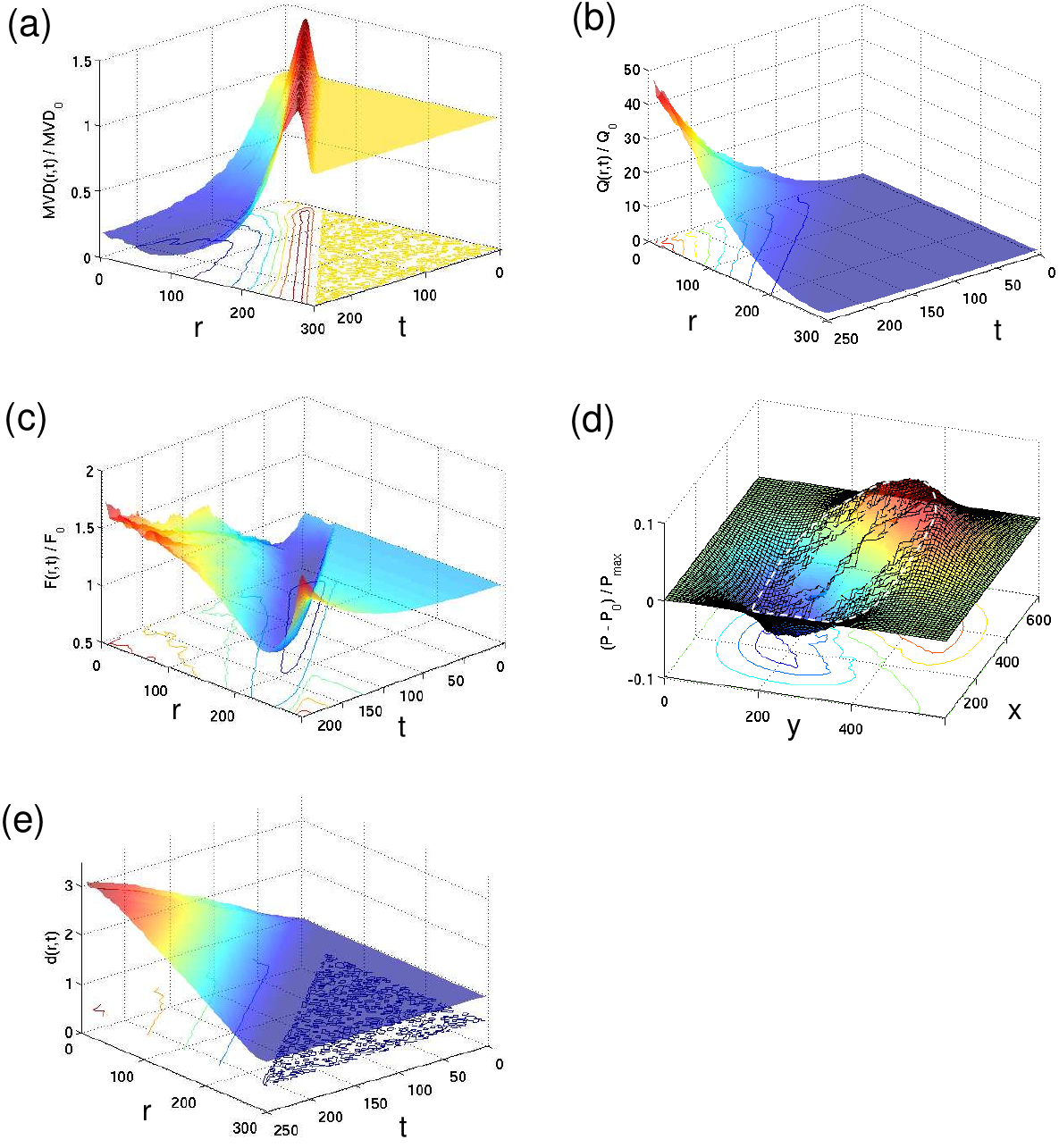}
}
    \caption{(Color online) As a function of radial distance
    from the tumor center, we plot in (a) normalized microvascular density
    MVD($r,t$)/MVD$_0$, (b) normalized blood flow $Q(r,t)/Q_0$ per vessel,
    (c) normalized shear force $F(r,t)/F_0$ on the vessel wall, (e) the
    average vessel radius $d(r,t)$.  Subplot (d) displays the difference in
    blood pressure field $P(x,y)$ between times $t=$ 150 and 0. The tumor is
    enclosed by the dotted region.} 
  \label{main_plot_nonuniform_vessel}
\end{figure*}
In Fig.~\ref{main_plot_nonuniform_vessel}(a) we quantify micro vascular
density. Normalized MVD shows a sharp maxima at a distance
$r=R_\mathrm{MVD}(t)$ from the center of the tumor, which is 
approximately equal to the tumor radius $R_T$, and then decays rapidly
towards the center of the tumor. This scenario is in accordance with the
results reported in~\cite{Lee06}, but contradicts slow decay of MVD reported
in~\cite{Bartha06}.  

The normalized blood flow, presented in
Fig.~\ref{main_plot_nonuniform_vessel}(b), increases very rapidly towards the
center of the tumor, supports prior findings~\cite{Bartha06}. Since the
vessel radius $d(e)$ increases linearly towards the tumor center, the flow
increases as the fourth power (as per Eq.~\ref{eq_flow}) of the radius. Large
increase in the flow, does not allow us to see the minor fluctuations
(e.g., the dip at $r=R_\mathrm{MVD}$ in the previous
case~\ref{main_plot_uniform_vessel}(b)) in the current profile. 

Fig.~\ref{main_plot_nonuniform_vessel}(c) shows the normalized shear force as a
function of $r$ and $t$. We observe a sharp minima along the periphery of the 
tumor, where MVD displays a maxima. According to Eq.~\ref{eq_shear}, shear force is
proportional to vessel-radius $d(e)$ and the pressure gradient across
it. Since $d(e)=1$ remains constant outside the tumor, new vessel originated
due to angiogenesis decrease the pressure gradient which in turn drops the
shear force in the region of interest. However, inside the tumor increase of
$d(e)$ and pressure gradient causes an increase in the shear force. Our result
is in agreement with~\cite{Bartha06}, but disagrees with the sharp fall observed
near the center of the tumor, reported in~\cite{Lee06}.  

The blood pressure difference at a particular instant $t=150$ normalized by
the maximum pressure at $x=0, y=0$, is shown in
Fig.~\ref{main_plot_nonuniform_vessel} (d). The characteristic behavior of
$(P-P_0)/P_{max}$ resembles to the previous case with uniform vessel radius
and the results obtained in~\cite{Bartha06}. Far from the tumor periphery, the
pressure difference is zero, which decreases further as the blood flow entering
the tumor through the highly vascularized peritumoral region. To preserve the
flow conservation, the pressure gradient increases inside the tumor along
the flow direction. 

The average vessel radius is constant outside the tumor. It increases almost
linearly with time, once we move from the periphery towards the tumor center,
as shown in the Fig.~\ref{main_plot_nonuniform_vessel} (e). In the vicinity near
the center of the tumor, linear increase in the vessel radius is slowed down
and tend to saturate toward the maximum value, although the maximum vessel
radius is never reached by the system. Results reported
in~\cite{Bartha06,Lee06} also shows similar profile.   

\section{Pressure correlated vessel collapse does not lead to realistic tumor
  vasculature}
\label{appC}
\begin{figure*}[h]
  \resizebox{2.0\columnwidth}{!}{%
    \includegraphics{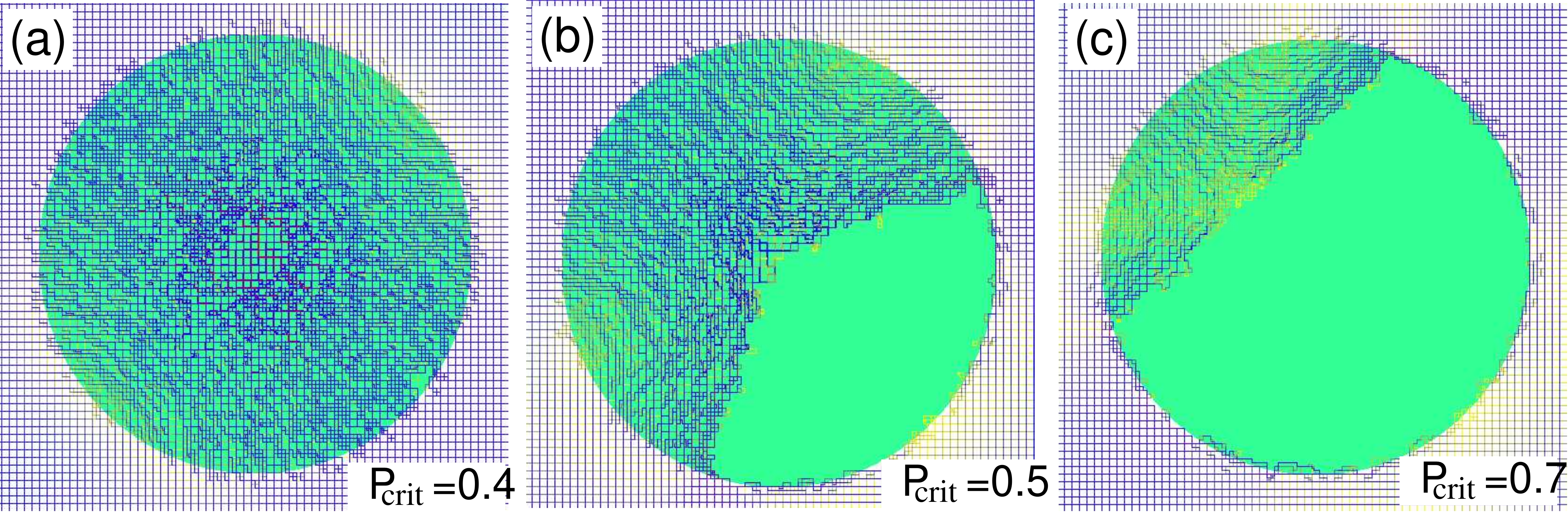}
  }
  \caption{(Color online) Snapshots of tumor vasculature at time $t=200$ for pressure
    correlated vessel collapse. Small (a)  and medium (b) values of critical pressures
    ($P_{crit}=$0.4, 0.5) show percolating vasculature and large (c) value of
    critical pressure ($P_{crit}=0.7$) leads to nearly vanishing vasculature inside
    the tumor. The color code is the same as described in
    Fig.~\ref{Pinf_snpshot_vesDia}.}  
  \label{press_vessel_collapse}
\end{figure*}
 Like shear force, blood pressure could also be a potential candidate for the
 flow-correlated percolation. Here we discuss the morphology of the tumor
 vasculature emerged under vessel collapsed below a normalized critical
 pressure $P_{crit}$. Simulation results for different $P_{crit}$ are shown in
 Fig.~\ref{press_vessel_collapse}. For $P_{crit}=$0.4, 0.5 we see a percolating vasculature
 perpendicular to the flow direction. Similar percolation ceases to occur for
 higher values of $P_{crit}(>0.7)$. Moreover, under the pressure correlated
 vessel collapse, the resulting tumor vasculaure appears entirely different
 from that we have seen in the main section of this paper.

\section{Tumor vasculature is not affected by the angiogenic regime} 
\label{appD}
\begin{figure*}[h]
  \resizebox{2.0\columnwidth}{!}{%
    \includegraphics{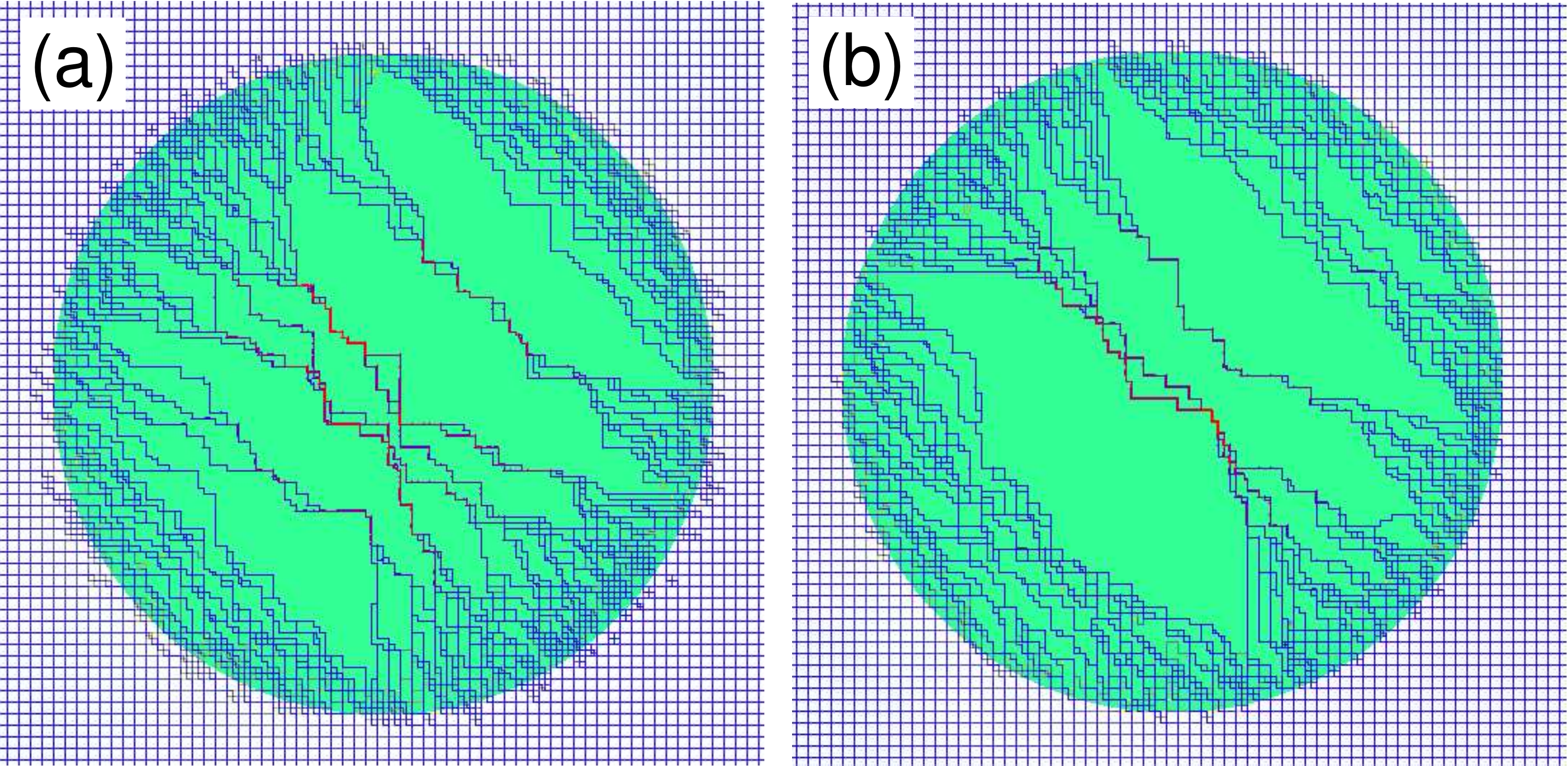}
  }
  \caption{(Color online) Snapshots of tumor vasculature at time $t=200$ for
    angiogenesis occurring (a) outside, (b) both inside and outside the tumor
    periphery. The color code is the same as described in 
    Fig.~\ref{Pinf_snpshot_vesDia}.}  
  \label{delta_angio_vessel}
\end{figure*}
We have simulated our model by translating angiogenic regime inside and outside
the tumor periphery. Earlier, in this paper, we have considered angiogenesis
occurring only outside the tumor periphery. Here we compare our prior findings
with the results obtained from the vessel proliferation within the annular
ring extending from $R_\mathrm{T}-\Delta_{\mathrm{angio}}/2$ to
$R_\mathrm{T}+\Delta_{\mathrm{angio}}/2$.  Our result, shown in Fig.~\ref{delta_angio_vessel},
suggest that the choice of $\Delta_{\mathrm{angio}}$ does not affect the tumor
morphology in large time limit.  

\section{Aging of tumor vessel enhances the percolation} 
\label{appE}
\begin{figure*}[h]
  \resizebox{2.0\columnwidth}{!}{%
    \includegraphics{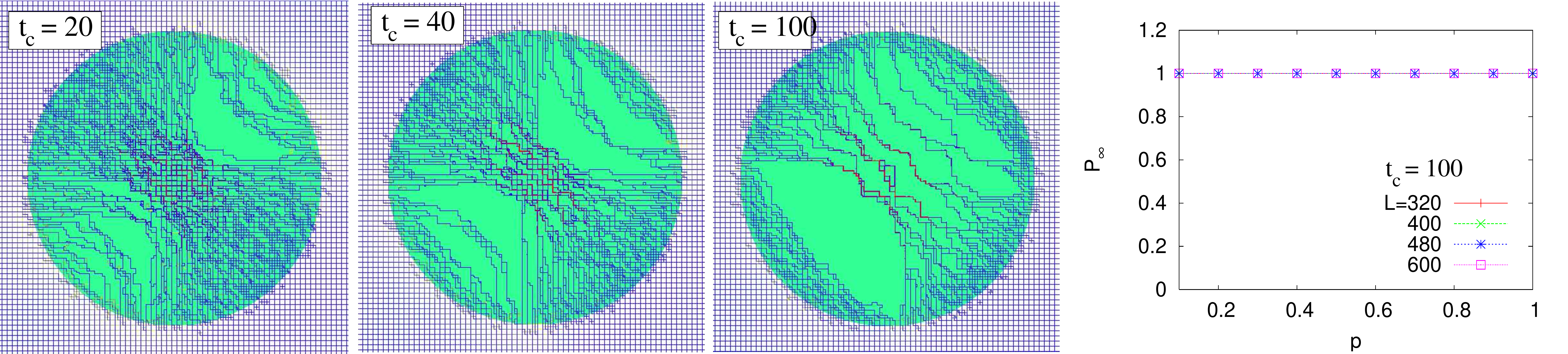}
  }
  \caption{(Color online) Snapshots of tumor vasculature at time $t=200$ with
    aging and shear force correlated vessel collapse. Morphologies 
    correspond to different values of $t_c$ (20,40,100) and a fixed
    $F_c=0.5$. The color code is the same as described in 
    Fig.~\ref{Pinf_snpshot_vesDia}. In the right most subplot $P_{\infty}$ is
    shown as a function of collapse probability $p$ for $t_c=100$ with system
    sizes $L=$320, 400, 480 and 600.}  
  \label{aging_shear_collapse}
\end{figure*} 
Earlier, in Fig.~\ref{Pinf_snpshot_without_angio}, we have seen that aging of
vessels inside the tumor 
works in favor of the percolating vasculature: smaller the $t_c$ (quicker the
vessels stabilize against collapse), greater the percolation
threshold. However, for any realistic values of $t_c$ vasculature inside tumor
ceases to exist at high collapse probabilities. On the contrary, in
Fig.~\ref{Pinf_snpshot_vesDia}, we have seen that, shear correlated vasculature is  
ever percolating. Here we study the effect of aging on shear correlated
collapse. Our results are depicted in Fig.~\ref{aging_shear_collapse}. At small $t_c$ we see a
dense vasculature inside the tumor which slowly reduces to normal shear
correlated vasculature at large $t_c$. It is
clear from the snapshots and percolation studies for different value of $t_c$, that both
aging and shear force work together to enhance the stability of the
tumor vasculature. 

\end{appendix}


\end{document}